\documentclass[useAMS,usenatbib]{mn2e}
\usepackage{mathtools}
\usepackage{graphicx}
\usepackage{color}
\usepackage{amsmath}
\usepackage{amssymb}
\usepackage{subfig}

\def\h2{H$_2$}

\usepackage[breaklinks,colorlinks,
  urlcolor=blue,citecolor=blue,linkcolor=blue]{hyperref}

\def\MH2c{\, M_{\rm H_2,cell}}
\def\Msunpc2{\,\rm M_{\odot}\,pc^{-2}}

\title[Spin alignment in stellar clusters]{The alignment is in their stars: on the spin-alignment of stars in star clusters}

\author[Ramon Rey-Raposo et al.]{Ramon Rey-Raposo$^{1}$\thanks{E-mail:
moncho.rey@gmail.com;} and Justin I. Read$^{1}$\\
$^{1}$Faculty of Engineering and Physical Sciences, University of Surrey, Guildford, GU2 7XH\\}
\begin{document}

\date{March 2018}

\pagerange{\pageref{firstpage}--\pageref{lastpage}} \pubyear{2017}

\maketitle

\label{firstpage}

\begin{abstract}
We simulate star formation in two molecular clouds extracted from a larger disc-galaxy simulation with a spatial resolution of $\sim 0.1$\,pc, one exiting a spiral arm dominated by compression, and another in an inter-arm region more strongly affected by galactic shear. Treating the stars as `sink particles', we track their birth angular momentum, and the later evolution of their angular momentum due to gas accretion. We find that in both clouds, the sinks have spin vectors that are aligned with one another, and with the global angular momentum vector of the star cluster. This alignment is present at birth, but enhanced by later gas accretion. In the compressive cloud, the sink-spins remain aligned with the gas for at least a free fall time. By contrast, in the shear cloud, the increased turbulent mixing causes the sinks to rapidly misalign with their birth cloud on approximately a gas free-fall time. In spite of this, both clouds show a strong alignment of sink-spins at the end of our simulations, independently of environment.
\end{abstract}


\section{Introduction}

The densest parts of molecular clouds -- molecular cloud cores  --  collapse under their own gravity, protected from any external radiation by an envelope of dust. These are the sites where new stars are born \citep{andre2013}. However, the fine details of this process of star formation remain unclear. One potentially powerful constraint on formation channels comes from the alignment or non-alignment of star spins inside star clusters. The latest theoretical models find that the angular momentum of molecular cloud cores depends primarily on small scale turbulence, though breaking due to magnetic fields that can also reduce their angular momenta \citep{2007ARA&A..45..565M}. The spin-alignment of stars in stars clusters depends, therefore, on what fraction of the kinetic energy of the host cloud is in rotational versus turbulent pressure support \citep[e.g.][]{Corsaro2017}. 
The observational picture of spin alignment, however, remains rather murky. Early work from \cite{Jackson&Jeffries2010} used estimates of the radii of stars, combined with spectroscopic rotation measurements, to estimate spin-alignment in the Pleiades and Alpha Per star clusters (with masses, $\sim800$\,M$_\odot$ and $\sim 350$\,M$_\odot$, respectively). They found no evidence for strong alignment, a result that persists even given new and larger estimates of the radii of M dwarf stars in Pleiades \citep{2018MNRAS.476.3245J}. By contrast, \cite{2018A&A...612L...2K} recently used a similar technique to rule out an isotropic distribution of spins in the Praesepe cluster (with stellar mass 500-600\,M$_\odot$), but see \citealt{2018MNRAS.tmp.1301K} for a discussion of some of the difficulties inherent in this measurement. Finally, using a different technique that measures stellar spins using astroseismology, \citet{Corsaro2017} found that stars in the open clusters NGC 6791 and NGC 6819 (with stellar masses $\sim$5000 M$_{\odot}$ and $\sim$2600 M$_{\odot}$, respectively) appear to have highly aligned spin vectors. Using  simulations with a spatial resolution of 0.002\,pc \citep[see][]{Lee&Hennebelle2016}, they show that such high alignment may be achieved if $>50$\% of the kinetic energy of the natal cloud is in rotation. (Note that all of these studies can be compatible if higher mass star clusters show stronger spin-alignment than lower mass clusters, or if astroseismology is a more reliable probe of spin-alignment than spectro-photometric techniques.)

As discussed above, from a theoretical perspective the spin-alignment of stars in star clusters depends sensitively on the turbulent velocity field of their birth clouds. This, in turn, depends on the galactic environment of the cloud \citep{Renaud2013,rrr2015}. Gas clouds formed in spiral arms are shocked and compressed, leading to the formation of more stars, while if a cloud forms in an inter-arm region, it is more dominated by shear, with less star formation. In this Letter, we study whether molecular clouds extracted from a spiral (compressive) or inter-spiral (shear) region lead to systematic differences in the spin-alignment of their stars. We run our simulations at a resolution of 0.1 pc, which is around 2 orders of magnitude smaller than the half light radius of 2 typical open clusters NGC 6791 and NGC 6819 (10\,pc and 7\,pc, respectively). We use sink particles to model the formation of stars. Our goal is to determine whether spin-alignments are ubiquitous, given realistic initial conditions for the birth-cloud, or whether they depend on environment.
\section{Details of the Simulations}
\begin{figure*} 
\includegraphics[width=170mm]{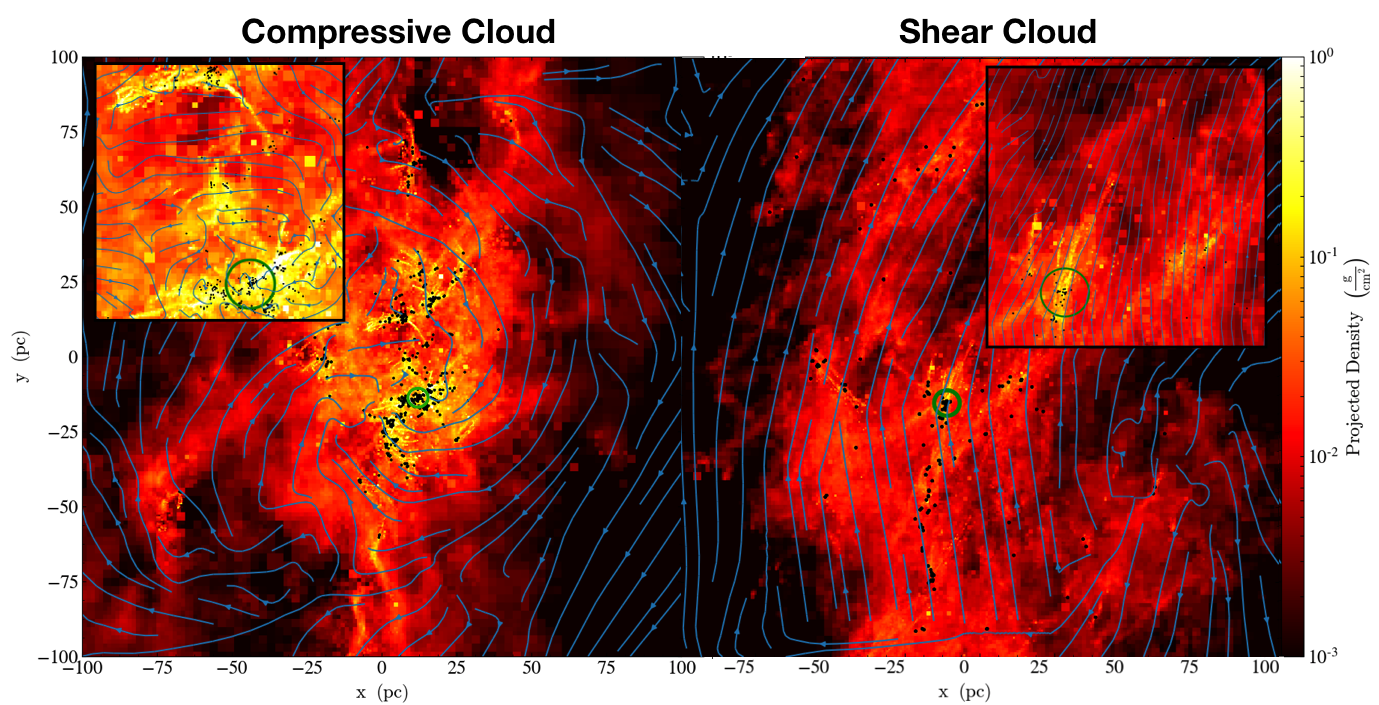}
\caption{Column density of the gas at 1 t$_{\rm ff}$, with the velocity field overlaid using blue lines. The dark dots represent the sinks and the green circle is the radius of the identified cluster. We have also included two panels with a zoom in map of the region around the cluster.} 
\label{fig:column}{}
\end{figure*}
For this work, we use the same initial clouds as in \citet{rrr2017}, using a resolution of $\sim$0.1 pc that is sufficient to resolve filaments and molecular cores \citep{arzoumanian2011characterizing,konyves2015census}. These clouds are a high-resolution version of the ones formed in galactic simulations \citep{Dobbs2013a,Dobbs2015}. We select all the particles included in a rectangular volume in the galactic models and increase their resolution \citep[following the procedure in][]{rrr2015}. The purpose of this Letter is to study the coupling of the angular momentum between the cloud and the molecular cores at an early stage of their formation before stellar feedback impacts further star formation. As such, we do not include a model for stellar winds or supernovae. Our sink particles represent molecular aggregates of a few hundred solar masses with a typical radius of 0.1 pc. For this study we consider two cases: Cloud B, which is an example of a cloud undergoing collapse as it has passed through a spiral arm and with mass of 2.6 $\cdot 10^6$ M$_{\odot}$; and Cloud C, whose evolution is characterised by the effect of the galactic shear stretching the cloud in the inter-arm space, and with mass of 1.4 $\cdot 10^6$ M$_{\odot}$. Both clouds have an approximate radius of 100 pc.

We use the modified version of \textsc{gadget2} \citep{2005MNRAS.364.1105S} used in \citet{rrr2017}. We have modified the sinks subroutine from \citet{Jappsen2005} to keep track of the initial angular momentum and its modification due to accretion of gas. The code also includes cooling and heating from \citet{Glover2007a,Glover2007b} and H$_2$ and CO chemistry (following \citet{Bergin2004} and \citet{Nelson1997}). The sink particles have a threshold density of $\rho_{\rm sink} = 1.6 \times 10^4$ cm$^{-3}$ with a sink radius R$_{\rm sink}$ = 0.1 pc. These parameters ensure that the minimum mass of a sink is over the Jeans Mass for each cloud, with a temperature of 50\,K (the temperature at which we initialise the clouds). We run each cloud for 1 free fall time (t$_{\rm ff}$) corresponding to 2.3\,Myr for the compressive cloud and 4.0\,Myr for the shear cloud. 

When a sink is created, the net angular momentum of the gas particles that form the sink is stored in a vector $\vec{J}^o_j$, where the superscript `$o$' refers to the creation of the sink and $j$ is the sink number in the cloud. As the sink accretes gas it modifies its angular momentum, so $\vec{J}_j(t)$ represents the angular momentum of the sink at a given time. We calculate the global angular momentum, applying:
\begin{equation}
\vec{J} = \sum_i \vec{r}_i \times m_i \vec{v}_i,
\label{eq:Jcloud}
\end{equation}
to the \textsc{sph} gas particles (if referring to the gas), or to the individual sinks (if referring to the star cluster). We also define an extra quantity:
\begin{equation}
\vec{L} = \frac{\vec{J}}{|\vec{J}|},
\label{eq:Lcloud}
\end{equation}
which is the direction of the angular momentum of the cloud.\\

Therefore it follows that if a sink is aligned with a given angular momentum, $\vec{L} \cdot \vec{L}_j$ = 1.  Thus, we can define the alignment of the sinks with their clouds, with each other, or with their host star cluster as:
\begin{equation}
A_j =  \vec{L}_i \cdot \vec{L}_j.
\label{eq:Aj}
\end{equation}
where $\vec{L}_i$ refers to the angular momentum of the birth gas cloud, the host star cluster, or the mean angular momentum of the sinks, depending on which alignment we wish to measure.

\section{Results}
In Fig. \ref{fig:column} we show the gas column density map for both clouds, giving a visual impression of their different structure. The sink particles are marked by the black dots, while the centre and radius of the most massive star cluster in each cloud is marked with a green circle. The projection of the direction of the velocity field is shown with blue arrows. We include two zoom in panels of the region around the most massive star clusters.

\begin{figure} 
\includegraphics[width=90mm]{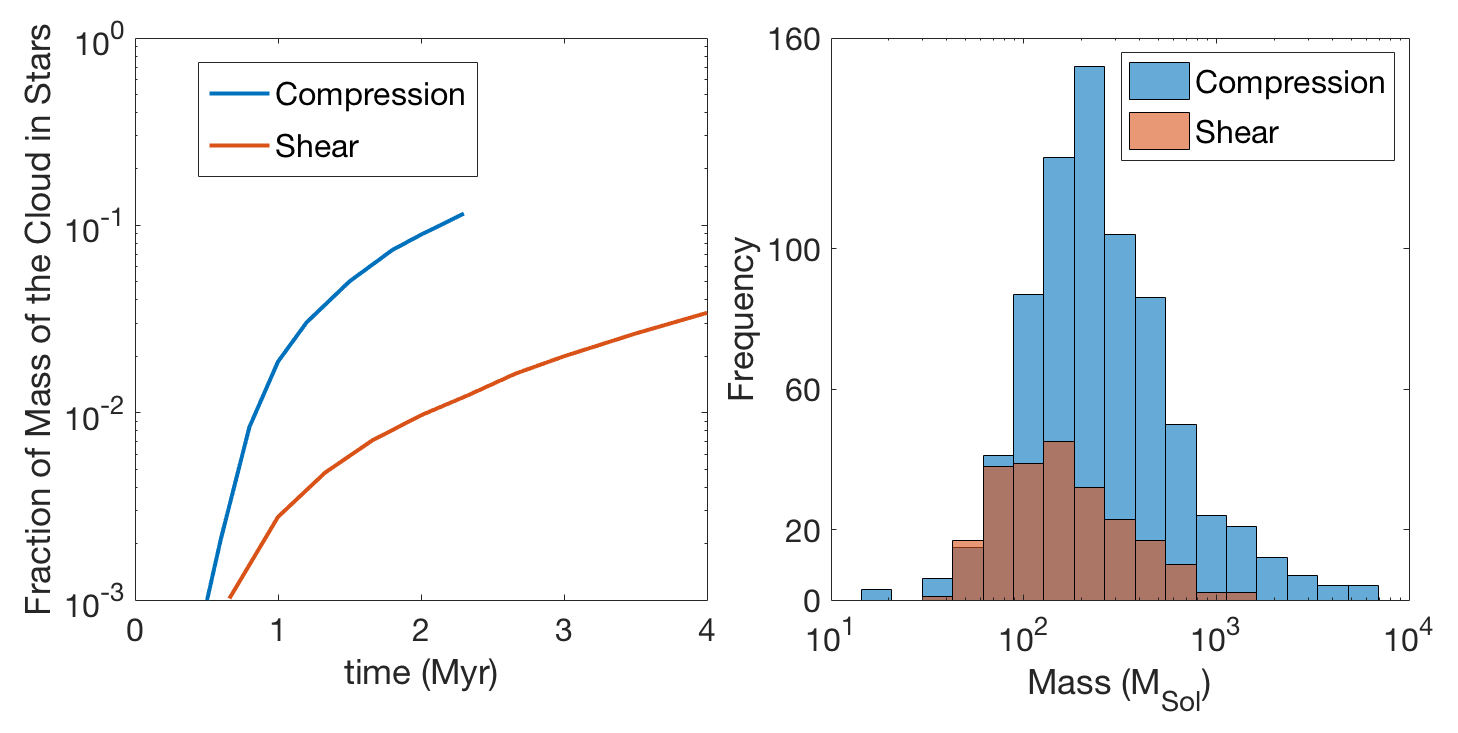}
\caption{The left panel shows the fraction of mass contained in sinks for both clouds. In the right panel we present the histograms for the distribution of masses of the sinks. The average sink mass is higher for the compressive cloud and it shows a longer tail towards the high mass end.}
\label{fig:SFR}{}
\end{figure}
As described in \cite{rrr2015} and \cite{rrr2017}, the compressive cloud forms stars at a higher rate and with higher masses. In the left panel of Fig. \ref{fig:SFR} we show the evolution of the fraction of mass contained in sinks for each cloud. At 1 t$_{\rm ff}$ the compressive cloud has more than 10\% of its mass in sinks, whereas for the shear cloud, this fraction only reaches 3\%. The distribution of masses of the sinks in the clouds is also different. Even though the initial mass of the sinks in each cloud is similar, the velocity field of the compressive cloud promotes the accretion of gas by the sinks, with some of them reaching $\sim$5000\,M$_{\odot}$. Sinks are born with their angular momentum aligned to the local conditions of the gas (a sphere of radius equal to 2 R$_{\rm sink}$ = 0.2\,pc). This angular momentum may only be changed afterwards by the accretion of new gas particles. 
\begin{table*}
\label{tab:ch_2summary}
\caption{Total mass of gas and sinks at 1 t$_{\rm ff}$ for the cloud. The number of sinks, total mass in gas and in sinks and the Plummer radius for the most massive star cluster in each cloud.}   
\centering 
\label{tab:ch_2summary}
\begin{tabular}{c c c c c c} 

\hline\hline 

Cloud: & M$_{\rm gas}$ / M$_{\rm sinks}$ (M$_{\odot}$) &  &N$_{\rm sinks}$ in Cluster& M$_{\rm gas}$ / M$_{\rm sinks}$ (M$_{\odot}$) &  Radius of Cluster (pc) \\ 
\hline  

Compressive&2.30 $\times 10^6$ / 3.0 $\times 10^5$ & & 54 & $1.62\times 10^{4}$ /  $3.90 \times 10^{4}$  & 3.15 \\
Shear& 1.36 $\times 10^6$ / 3.7 $\times 10^4$& & 26 & $9.38\times 10^{3}$ / $6.16 \times 10^{3}$ & 4.26 \\
\hline 
\end{tabular} 
\end{table*}

This difference in the star formation process for the clouds is reflected also in the spatial distribution of the sinks. As shown in Fig. \ref{fig:column}, in the compressive cloud we can identify a few aggregates of stars in the centre of the cloud, each with a substantial number of sinks. This is not the case for the shear cloud, where sinks form in smaller quantities and they are more scattered. To locate the clusters in our clouds, we select the most massive cluster in each cloud. We find its density centre using a {\it shrinking sphere} method \citep{power2003}. We then calculate its radius by fitting a Plummer sphere \citep{plummer1911} to the density distribution of bound sinks. We define the `edge' of each star cluster to be twice its Plummer scale radius. Using this technique we select the most massive cluster for each cloud. The number and properties of star clusters found using this method are given in Table \ref{tab:ch_2summary}.

Although both clusters are similar in size ($\sim$ 3-4\,pc), the total number of sinks in the compressive cloud is more than double than for the shear cloud, while the mass enclosed is six times higher. This is in line with the results in \citet{rrr2017} and more recent simulations using the \textsc{ramses} code (Rey-Raposo et al. 2018, in prep.).

\begin{figure*} 
\includegraphics[width=140mm]{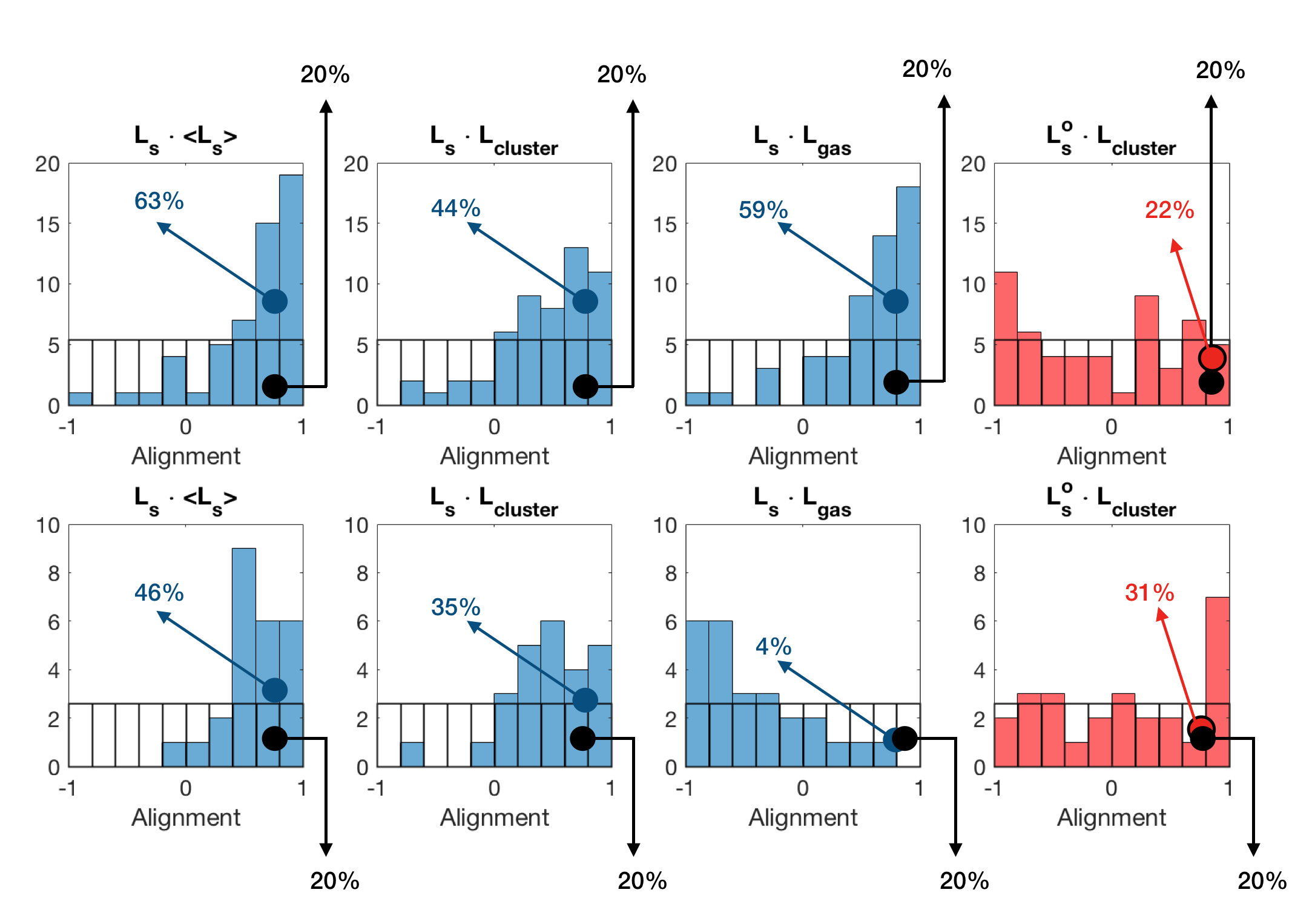}
\caption{Histogram of the alignment of the angular momenta for the two most massive star clusters in the clouds. The left column shows the alignment of the spin of the sinks with the average spin of sinks in the star cluster. The second column shows the alignment of the spin of the sinks with respect to the total angular momentum of the star cluster. The third column shows the alignment of the spin of the sinks with the remaining gas in the star cluster. The alignment of the original angular momentum of the sinks with the total angular momentum of the star cluster at 1 t$_{\rm ff}$ is shown in the last column. We also display the histogram of a uniform distribution of the same size in black for each panel. We show the percentage of sinks included in the last two bins on top of each histogram, for the data and the random distribution.} 
\label{fig:alignment}{}
\end{figure*}

For the sinks included in these clusters, we compute the directions of their average internal angular momentum, $<\vec{L}_s>$, their total angular momentum with respect to their host star cluster, $\vec{L}_c$, and the angular momentum of the gas in the host cloud $\vec{L}_{\rm gas}$ (using Eqs. \ref{eq:Jcloud} and \ref{eq:Lcloud}). In Fig.\ref{fig:alignment}, we show the histogram of alignment (defined as in Eq.\ref{eq:Aj}) of these quantities in both clouds. The alignment of the angular momentum of the sinks with the average of the angular momenta of all the sinks in the cluster is shown in the left panels of Fig. \ref{fig:alignment}. The second column presents the alignment of the sinks with the global angular momentum of the sinks in the cluster, and in the right panels we display the alignment of the angular momentum of the sinks with the angular momentum of the remaining gas in the cluster. In the last column, we present the alignment of the angular momentum of the sinks at creation with the total angular momentum of the cluster at 1 t$_{\rm ff}$. For each panel we have added another histogram depicting a uniform distribution of the same size as the data, and in each plot, we display the percentage of the number of sinks included in the last two bins, as a quantitative measure of the alignment.

In both clouds, we find a distribution of sinks with a tail to masses over 200\,M$_{\odot}$, representing very massive star forming clumps. We verified that these massive sinks do not bias our angular momentum results by showing that the histograms in Fig. \ref{fig:alignment} are unchanged if we remove all sinks with masses $>200$\,M$_{\odot}$. This insensitivity to the most massive sinks owes to the fact that they are formed preferentially near the centres of the star clusters where the gas density is highest.

\begin{figure} 
\includegraphics[width=80mm]{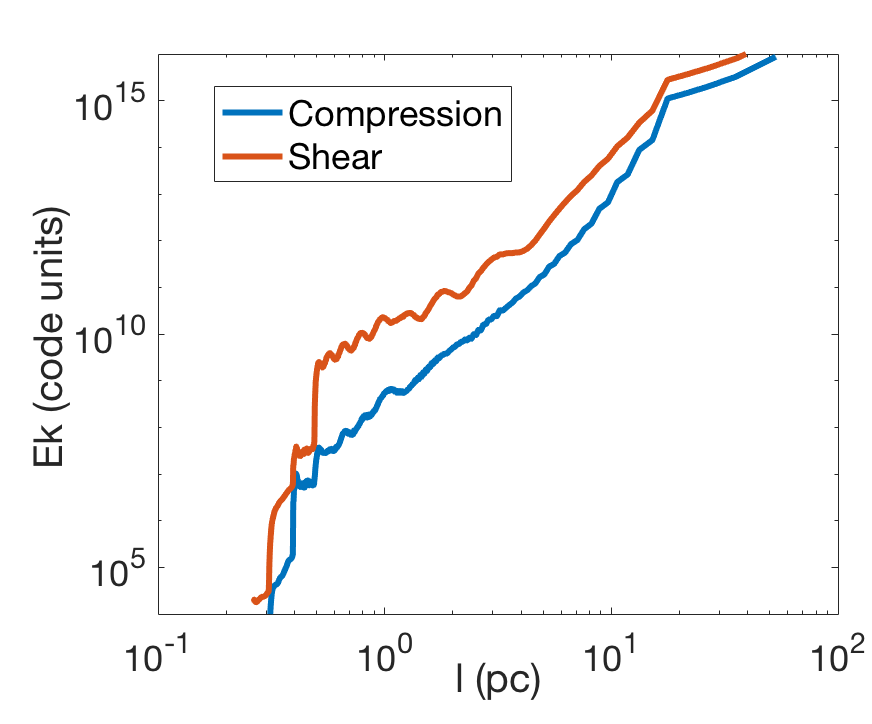}
\caption{Kinetic power spectrum of the two clouds expressed in code units. The $x$-axis shows the scale $l$ defined as $ l = 2 \pi /k$, where $k$ is the wavenumber. The shear cloud is more turbulent in general, and almost two orders of magnitude larger for scales around 1\,pc.} 
\label{fig:PS}{}
\end{figure}

For both clouds, the alignment of the initial spin of each sink with the global angular momentum of the cloud at 1 t$_{\rm ff}$ is, at most, weak. Sinks are created continuously after $\sim$ 0.1\,Myr, and, therefore, their initial spins reflect the angular momentum of the gas at creation. As the clouds evolve, new sinks are created and the existing ones accrete more gas, modifying their angular momenta. As a consequence there is a better alignment with the total angular momentum of the cluster ($J_{c}$). Sinks in the compressive cloud are slightly more aligned, as accretion is enhanced by the compressive velocity field. For both clouds we also find alignment between the sinks and the average angular momentum of the cluster. Lastly, we find strong alignment between the sinks and the angular momentum of the gas (in the region defined by the cluster) for the compressive cloud. This is not the case for the shear cloud, suggesting that the velocity field in the compressive cloud is more similar at both larger (R$_{clus} \sim$5 pc) and smaller scales (R$_{sink}$), whereas for the shear cloud there is some turbulent mixing on small scales.
To measure the effect of shear disrupting the alignment of the stars, we calculate the power spectrum following the method described in \citet{Grisdale+2018}. We select a box of size 50\,pc in the centre of each cloud at the beginning of each simulation. We present the kinetic energy power spectrum of the two clouds in Fig. \ref{fig:PS}, and we show the length scale $l = 2 \pi/k$ where $k$ is the wavenumber. The shear cloud possesses more kinetic energy at every scale, but the difference is more visible for lengths of the order of $\sim$0.5\,pc, where the turbulence is around two orders of magnitude higher for the shear cloud than for the compressive cloud, and where the local velocity field has an effect over the angular momentum of a newly created sink.

\section{Discussion}
Our clouds represent only two extreme cases of galactic environments; in reality, there should be a wide variety of intermediate situations. We find that turbulent motions erase the alignment between the gas and the star cluster in the shear cloud. However, in both clouds the stars' spins are aligned both with the global angular momentum of the star cluster and with each other. This is in accordance with the theoretical and observational results by by \cite{Corsaro2017} who find strong alignment in the open clusters NGC 6791 and NGC 6819, and in their simulations (if more than 50\% of the initial energy is in rotation).  In our case, the compressive and shear clouds have 41\% and 37\% of their initial kinetic energy in rotation modes, respectively. This is in reasonably good agreement with the 50\% bound found by \citet{Corsaro2017}, however we note that the comparison is complicated by the fact that our initial conditions represent a more evolved state for the clouds than in their simulations. The alignment of the initial angular momentum of the sinks with their average angular momentum, and with the total angular momentum of the cluster, suggests that the alignment of the sinks within themselves is enhanced via the continuous accretion of the gas.

Feedback affects differently the clouds \citep{rrr2017}: In the cloud dominated by shear, the strength of the galactic velocity field is $\sim10$ times higher than the impact of the feedback. This inherited turbulence at sub-pc scales causes the gas to misalign with the angular momentum vector of the sinks (see Figure \ref{fig:alignment}, right panel). Therefore, feedback may limit the turbulence mixing happening at sub-pc scales preventing the disalignment of sinks in the shear cloud. On the other hand, the effect of feedback in a very compressive environment is to reduce the star formation rate, by delaying the collapse of the gas and therefore the accretion, effectively freezing the angular momenta of the sinks. In both cases, however, it is unlikely that such effects would be strong enough to fully erase the alignments that we find here.

One of the major limitations of this work is our inability to resolve structures on scales smaller than the sink particles. This is why we refer to them as `molecular clumps' and not stars. The internal distribution of stars within these cores is not resolved in our simulations, and therefore changes in its internal angular momentum due to tidal and other effects are neglected.

Finally, our simulations neglect the role of magnetic fields. These will cause angular momentum loss through `breaking' \citep[e.g.][]{2007ARA&A..45..565M}, but are not likely to fully erase alignment. More troublesome may the role of star cluster mergers, however. If star clusters assemble through a large number of near equal-mass mergers, then the alignment signal will be erased. If however, star clusters are dominated by one massive progenitor then the alignment will be somewhat weakened, but not erased, by mergers. We will discuss the hierarchical build-up of star clusters in more detail in a forthcoming paper.
\section{Summary and Conclusions}
In this Letter, we have studied the effect of the galactic environment in the transference of angular momentum between a parent gas cloud and its molecular cores. Our results suggest that, at creation, the spin of a sink particle follows the angular momentum of the gas in its local galactic environment. Even in two very different galactic environments, we find that the sinks are aligned both with the global angular momentum of the cluster, and with the average angular momentum of the stars. Our results require confirmation from simulations of a larger number of molecular clouds at higher resolution and including the physics of magnetic fields and stellar feedback. However, the fact that we find that star spins are strongly aligned in two very different galactic environments suggests that star-spin alignments may be ubiquitous. This has interesting implications for the formation of massive stellar binaries and their gravitational wave emission. Our results suggest that such binaries are much more likely to be spin-aligned, leading to an incorrect inference of their stellar remnant properties if non-aligned templates are used \citep[e.g.][]{2017JKPS...70..735C,2018JCAP...03..007C}.
\section{Acknowledgements}
The calculations for this Paper were performed on the `Eureka' supercomputer at the University of Surrey. RRR and JIR would like to acknowledge support from STFC consolidated grant ST/M000990/1.
\bibliographystyle{mn2e}
\bibliography{RRR}
\label{lastpage}
\end{document}